\documentclass[12pt]{article}

\usepackage{scicite}

\usepackage{times}

\usepackage[utf8]{inputenc}
\usepackage{graphicx}
\usepackage{fancyhdr}
\graphicspath{ {figures/} }
\usepackage{xcolor}
\usepackage{amsmath,amsthm,amssymb}
\usepackage{graphicx} 
\usepackage{tikz,tkz-tab}
\usepackage[
colorlinks=true,
urlcolor=blue,
linkcolor=blue,
citecolor=blue
]{hyperref}
\usepackage{cite}

\def\url#1{}
\newcommand{\bs}{\boldsymbol}

\newenvironment{sciabstract}{%
\begin{quote} \bf}
{\end{quote}}

\title{Production and characterization of a fragmented spinor Bose-Einstein condensate}
\author
{Bertrand Evrard$^\ast$, An Qu, Jean Dalibard, Fabrice Gerbier
\\
\normalsize{Laboratoire Kastler Brossel,  Coll\`ege de France,} \\
\normalsize{ CNRS, ENS-PSL University, Sorbonne Universit\'e, }\\
\normalsize{11 Place Marcelin Berthelot, 75005 Paris, France}\\
\\
\normalsize{$^\ast$To whom correspondence should be addressed; E-mail:  bertrand.evrard@lkb.ens.fr}
}
\date{}

\begin{document}
\baselineskip24pt

\maketitle 

%Abstracts of Research Articles and Reports should explain to the general reader why the research was done, what was found and why the results are important. They should start with some brief BACKGROUND information: a sentence giving a broad introduction to the field comprehensible to the general reader, and then a sentence of more detailed background specific to your study. This should be followed by an explanation of the OBJECTIVES/METHODS and then the RESULTS. The final sentence should outline the main CONCLUSIONS of the study, in terms that will be comprehensible to all our readers. The Abstract is distinct from the main body of the text, and thus should not be the only source of background information critical to understanding the manuscript. Please do not include citations or abbreviations in the Abstract. The abstract should be 125 words or less. For Perspectives and Policy Forums please include a one-sentence abstract.

\begin{sciabstract}  % Just 125 words
Understanding the ground state of many-body fluids is a central question of  statistical physics. Usually for weakly interacting Bose gases, most particles occupy the same state, corresponding to a Bose--Einstein condensate. However, another scenario may occur with the emergence of several, macroscopically populated single-particle states. The observation of such fragmented states remained elusive so far, due to their fragility to external perturbations. Here we produce a 3-fragment condensate for a spin 1 gas of $\sim 100$ atoms, with anti-ferromagnetic interactions and vanishing collective spin. Using a spin-resolved detection approaching single-atom resolution, we show that the reconstructed many-body state is quasi-pure, while one-body observables correspond to a mixed state. Our results highlight the interplay between symmetry and interaction to develop entanglement in a quantum system.
\end{sciabstract}

\newpage 

Bose--Einstein condensation (BEC) is a remarkable low-temperature phenomenon, in which a macroscopic fraction of the particles of a fluid accumulate in the single-particle state $\psi_0$ of lowest energy, even though there may be many almost degenerate states nearby. 
Repulsive interactions play a decisive role in stabilizing the resulting condensate and ensuring that only $\psi_0$ acquires a macroscopic population, a \emph{winner takes all} situation \cite{leggett2001,nozieres1995}. However, BEC is not the only possible scenario for a fluid of interacting bosons. 
A situation opposite to BEC, achieved for instance in the Mott-insulator regime \cite{Fisher:1989}, occurs when the single-particle ground level is massively degenerate. There, no single particle state acquires a macroscopic population even at zero temperature. 

A long-sought situation, intermediate between these two cases, corresponds to a fragmented condensate \cite{nozieres1982,mueller2006}. Here, the many-body ground state leads to macroscopic populations for a few single-particle states $\psi_0$, $\psi_1$, \ldots, $\psi_p$ simultaneously. More precisely, for one-particle observables the situation is equivalent to a statistical mixture of the $p$ single-particle states, but a measurement performed at the few- or $N$-body level can reveal the quantum entanglement between the system constituents. Theoretically, fragmentation often results from an interplay between interactions and symmetry \cite{mueller2006}. The $N$-body Hamiltonian and its ground state possess a given symmetry, but the BEC-like (mean-field) states $\psi^{\otimes N}$ that minimize the average energy must break this symmetry. Fragmentation and entanglement thus provide a way to restore the symmetry and lower the energy with respect to the mean-field prediction.

A paradigm model for a fragmented condensate consists in an assembly of spin 1 atoms, in which the external degrees of freedom of the particles have been frozen \cite{mueller2006,law1998,ho2000,castin2001,koashi2000,ashhab2002,barnett2010,desarlo2013}. The atoms all share the same spatial wave function and only their spin degree-of-freedom is relevant \cite{yi2002}. Assuming the interaction Hamiltonian
\begin{equation}
\hat H_s=\frac{U_s}{2N}\sum_{i,j=1}^N\hat{\bs s}_i\cdot \hat {\bs s}_j=\frac{U_s}{2N}\hat {\bs S}^2,
%\abel{eq:}
\end{equation}
where $\hat{\bs s}_i$ is the spin operator of atom $i$ and $\hat {\bs S}=\sum_i \hat{\bs s}_i$ is the collective spin operator, 
the ground state is expected to be fragmented for anti-ferromagnetic interactions, \emph{i.e.} for a positive coupling constant $U_s$. For an even number of bosons $N$, the many-body ground state $|\Phi\rangle$ of $\hat H$ is a collective singlet state ($S=0$). Since it is rotationally invariant, the single-particle density matrix  $\rho^{(1)}$ extracted from $|\Phi\rangle$ must be the same in any single-particle basis $|m\rangle_{\bs u}$, where $m=0,\pm 1$ is the magnetic quantum number and $\bs u$  the orientation of the  quantization axis. 
This can be achieved only if $\rho^{(1)}$ is proportional to the identity matrix $\hat 1$, with its three eigenvalues equal to $1/3$, thus providing a $p=3$ fragmented condensate. When $N$ is odd, $|\Phi\rangle$ is a state with collective spin $S=1$ and also corresponds to a 3-fragment condensate. 

Here we present the first experimental characterization of a fragmented condensate prepared close to the collective singlet state, using $N\approx 100$ sodium atoms tightly confined in an optical dipole trap. The only relevant degrees of freedom are the Zeeman states $m=0,\pm1$ of each atom, which can evolve through binary collisions $2\times|m=0\rangle \leftrightarrows |m=+1\rangle + |m=-1\rangle$. This process conserves the magnetization $\hat{S}_z=\hat{N}_{+1}^{(z)}-\hat{N}_{-1}^{(z)}$, where the operator $\hat{N}_m^{(z)}$ counts the number of atoms in state $m$, with $z$ as quantization axis. The relevant Hamiltonian in the presence of a static magnetic field $B$ parallel to $z$ reads 
	$\hat{H}=-q\hat{N}_0^{(z)}+\hat H_s\,.$ 
Here the linear Zeeman shift $\propto S_z$ is omitted since $S_z$ is a conserved quantity and we keep only the quadratic Zeeman shift, which lowers the energy of $m=0$ with respect to $m=\pm 1$ ($q\propto B^2$). For $q\gg U_s$, spin-spin interactions can be neglected and the ground state $|\Phi\rangle$ is a single, uncorrelated condensate in the state $|m=0\rangle_z^{\otimes N}$. As long as $q\gg U_s/N^2$, this $m=0$ condensate remains dominant and its small depletion can be evaluated using a standard Bogoliubov approach \cite{SM}. For a lower magnetic field $q\lesssim U_s/N^2$, the depletion becomes extensive and $|\Phi\rangle$ is fragmented. Finally when $q$ is strictly zero, $|\Phi\rangle$ is the  $S=0$ (resp. $S=1$) state mentioned above for even (resp. odd) $N$.  

\begin{figure}[h!]
	\centering
	\includegraphics[width=16cm]{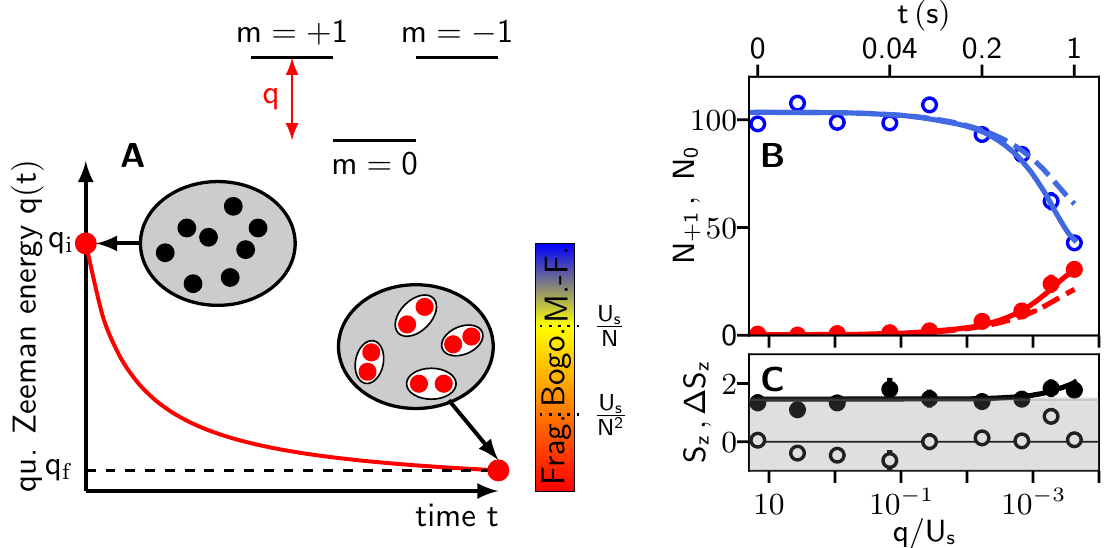}
	\caption{{\bf A.} Sketch of the adiabatic ramp used to produce a fragmented condensate. Initially in a large magnetic field ($q_i\gg U_s$), the system forms a single condensate with all atoms in $m=0$. When the magnetic field is decreased to a low value ($q_f\sim U_s/N^2$), the condensate fragments. For $N$ even, the ground state is close to the singlet state $S=0$, which can be viewed \cite{ho2000} as a condensate of $N/2$ pairs of atoms, each in the zero angular momentum state $|0,0\rangle-|+1,-1\rangle-|-1,+1\rangle$. For $N$ odd, the ground state is a condensate of such pairs, plus an extra single atom. {\bf B.} Evolution of the populations in $m=0$ (blue circles) and $m=+1$ (red dots) over the ramp. The solid and dashed lines show the predictions from the $N$-body Schr\"odinger equation, averaged over the parity of the initial state. For the solid line we added two stochastic elements to the Schr\"odinger equation (quantum trajectory method), to model the quantum jumps associated with one-atom loss and ``spin-flip" processes. {\bf C.} Evolution of the average magnetization (black circles) and its standard deviation (dots). Both remain at the level of the detection noise (gray area). Error bars show the statistical error corresponding to two standard deviations.}
\label{figure1}
\end{figure}

Because of the smallness of the energy gap ($<100\,$pK) protecting the many-body ground state in our system, a direct cooling to the fragmented state is not possible. Instead, we use an adiabatic passage induced by ramping down the magnetic field, which drives the system from the uncorrelated state at large $B$ to the targeted state at $B\approx 0$ (Figure \ref{figure1}{\bf A}). The populations in each Zeeman state are measured with a resolution of $1.2$ atom using a combination of Stern-Gerlach separation and fluorescence imaging \cite{qu2020}. Fig.\,\ref{figure1}{\bf B} gives the evolutions of the measured mean $N_0$ and $N_1$ over the ramp, and shows that the three Zeeman states end up with comparable populations. We also plot in Fig.\,\ref{figure1}{\bf B} the results of a numerical solution of the $N$-body Schr\"odinger equation, taking into account the residual decoherence during the ramp via additional stochastic elements \cite{SM}.  Importantly, the magnetization $S_z$ (Fig.\,\ref{figure1}{\bf C}) remains compatible with zero and its deviation $\Delta S_z=1.88\,(10)$ at the end of the ramp can be attributed mostly to detection noise. 

\begin{figure}[h!]
	\centering \includegraphics[height=5.5cm]{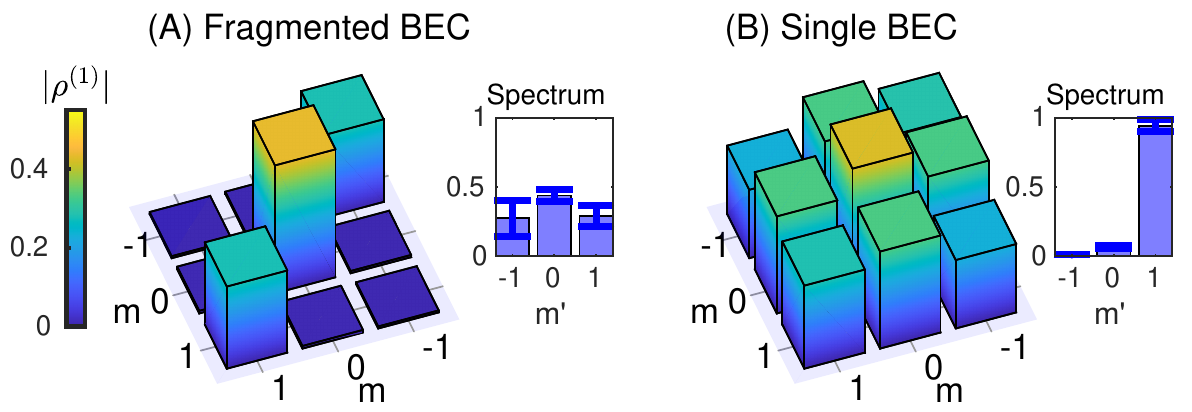}
	\caption{Modulus of the elements of the one-body density matrix and its eigenvalues for a fragmented BEC ({\bf A}) and for a single BEC ({\bf B}).}\label{figure_rho_1}
\end{figure}

We now investigate the state of the system at the end of the ramp, focusing successively on one-, two- and $N$-particle observables. Our main experimental tool consists in rotations in spin space around $z$ and $y$ axes with adjustable angles, followed by imaging. Rotations around $z$ are obtained from Larmor precession in the static magnetic field, and rotations around $y$ are induced by a resonant radio-frequency field \cite{SM}. Consider first the reduced one-body density matrix $\rho^{(1)}_{m,m'}=\langle \hat a_{m'}^\dagger \hat a_m\rangle$, where $\hat a_m$ annihilates a particle in state $m$. Using linear algebra, the average values $N_m(\phi)$ measured after rotation around $z$ by a series of angles $\phi$, followed by a rotation $\pi/4$ around $y$, allow one to reconstruct all nine real coefficients of $\rho^{(1)}$, shown in Fig.~\ref{figure_rho_1}\,{\bf A}. We find that $\rho^{(1)}$ has similar diagonal elements and essentially zero off-diagonal ones. The three eigenvalues are thus comparable and the von Neumann entropy $\mathcal{S}(\rho^{(1)})=1.07^{+0.02}_{-0.10}$ is very close to the upper bound $\ln3\simeq1.10$ for a completely mixed state. As a control experiment, we also reconstructed $\rho^{(1)}$ for the  condensed state  $\propto \left(|-1\rangle_z+\sqrt 2|0\rangle_z+|1\rangle_z\right)^{\otimes N}$ (Fig\,\ref{figure_rho_1}\,{\bf B}). There, we find comparable diagonal and off-diagonal elements $\rho^{(1)}_{m,m'}$. After diagonalization, we find that one eigenvalue $\simeq 0.94\,(4)$ is dominating, as expected for a single condensate.

\begin{figure}[h!]
	\centering
	\includegraphics[width=12cm]{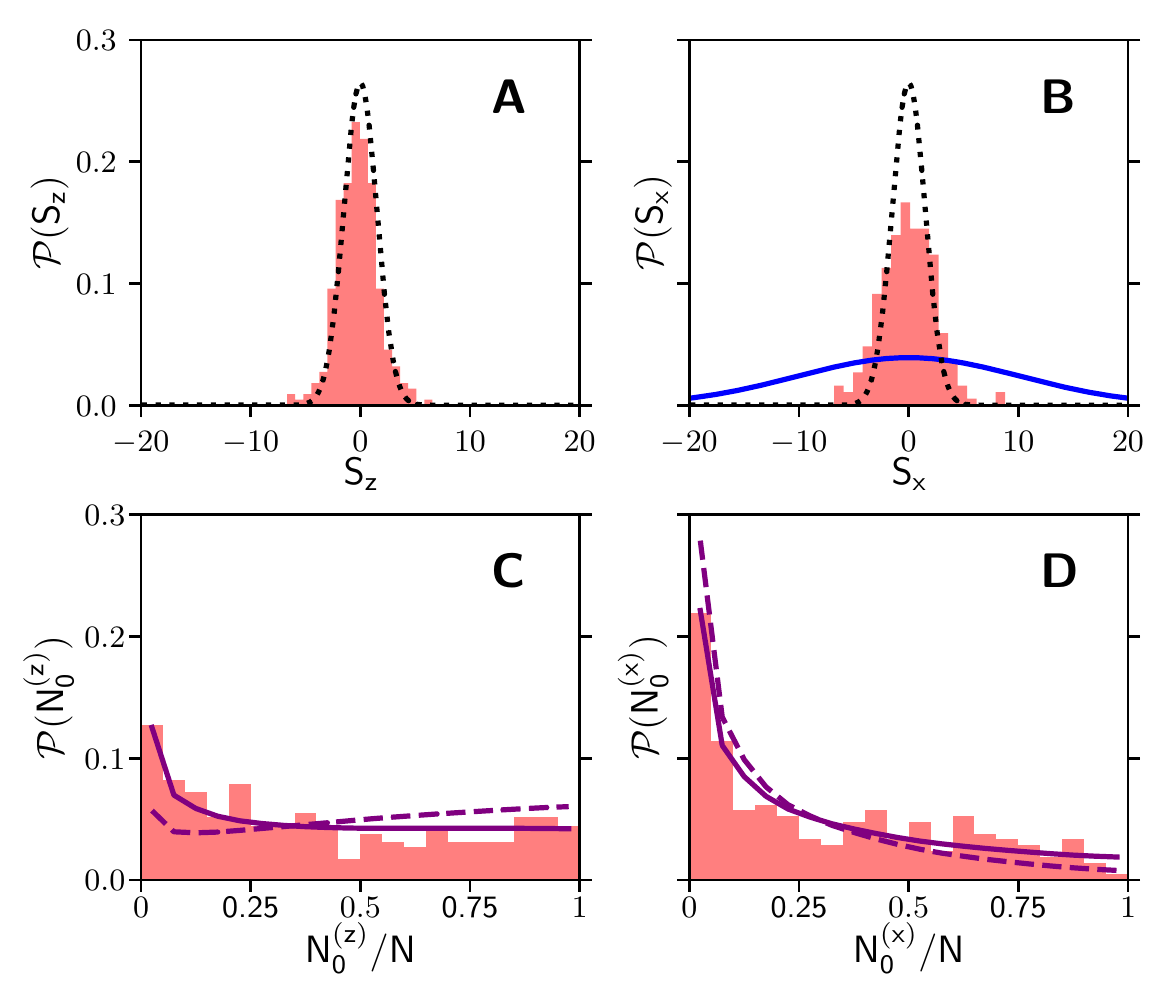}
	\caption{Top row: Distribution of results for $S_z=N^{(z)}_{+1}-N^{(z)}_{-1}$ ({\bf A}) and for $S_x$ ({\bf B}). Both distributions are very close to the detection noise (dotted line). For comparison, we also plot in ({\bf B}) the predicted distribution $\mathcal{P}(S_x)$ for the polar state $|m=0\rangle_z^{\otimes N}$ (blue line). Bottom row: Distribution of results for $N^{(z)}_{0}$ and $N^{(x)}_{0}$. As in Fig. \ref{figure1}B, the solid and dashed lines show the predictions from the $N$-body Schr\"odinger equation. For the dashed line, the equation is purely deterministic whereas for the solid line, two additional stochastic elements were added to mimic the residual decoherence processes during the preparation ramp.}
\label{fig:two_body}
\end{figure}

The fragmentation revealed by $\rho^{(1)}$ can be caused by either quantum or thermal fluctuations. Two-body observables such as $\hat{\bs S}^2$ provide a first information on the existence of quantum entanglement in the system. Here the value of $\langle S_z^2\rangle$ is calculated from the distribution ${\cal P}(S_z)$ of the results for $S_z$, given in Fig. \ref{fig:two_body}A. The values of $\langle S_i^2\rangle$, $i=x,y$, are obtained in a similar way after proper rotation of the state. We show the distribution ${\cal P}(S_x)$ in Fig. \ref{fig:two_body}B. For each spin component $i$, we find a highly squeezed distribution around $S_i=0$ and we find after deconvolution of the detection noise $\langle\hat{\bs S}^2\rangle= 9.9\,(1.0)$, which is $\sim 20$ smaller than the value $\langle\hat{\bs S}^2\rangle=2N$ expected for independent spins.
%Since the collective spin of $N$ independent atoms is larger than $\sqrt{2N}$, the squeezing parameter
%\begin{equation}
%	\xi_s=\frac{\langle\hat{\bs S}^2\rangle}{2N}
%	\label{eq:squeezing}
%\end{equation}
%detects entanglement when $\xi_s<1$. Here we find after deconvolution of the detection noise $\langle\hat{\bs S}^2\rangle= 9.9\,(1.0)$, thus $\xi_s\simeq 0.048$, $13\,$dB below the separability limit.  
The counterpart of this large spin squeezing is a very broad distribution of the populations of the three spin states. We show in  Fig. \ref{fig:two_body}C and D the distributions ${\cal P}(N_0^{(i)})$ along the $i=z,x$ axes. They are similar, as expected for an essentially isotropic state, and they are in agreement with the predictions from the solution of the $N$-body Schr\"odinger equation.  

\begin{figure}[h!]
	\centering
	\includegraphics[]{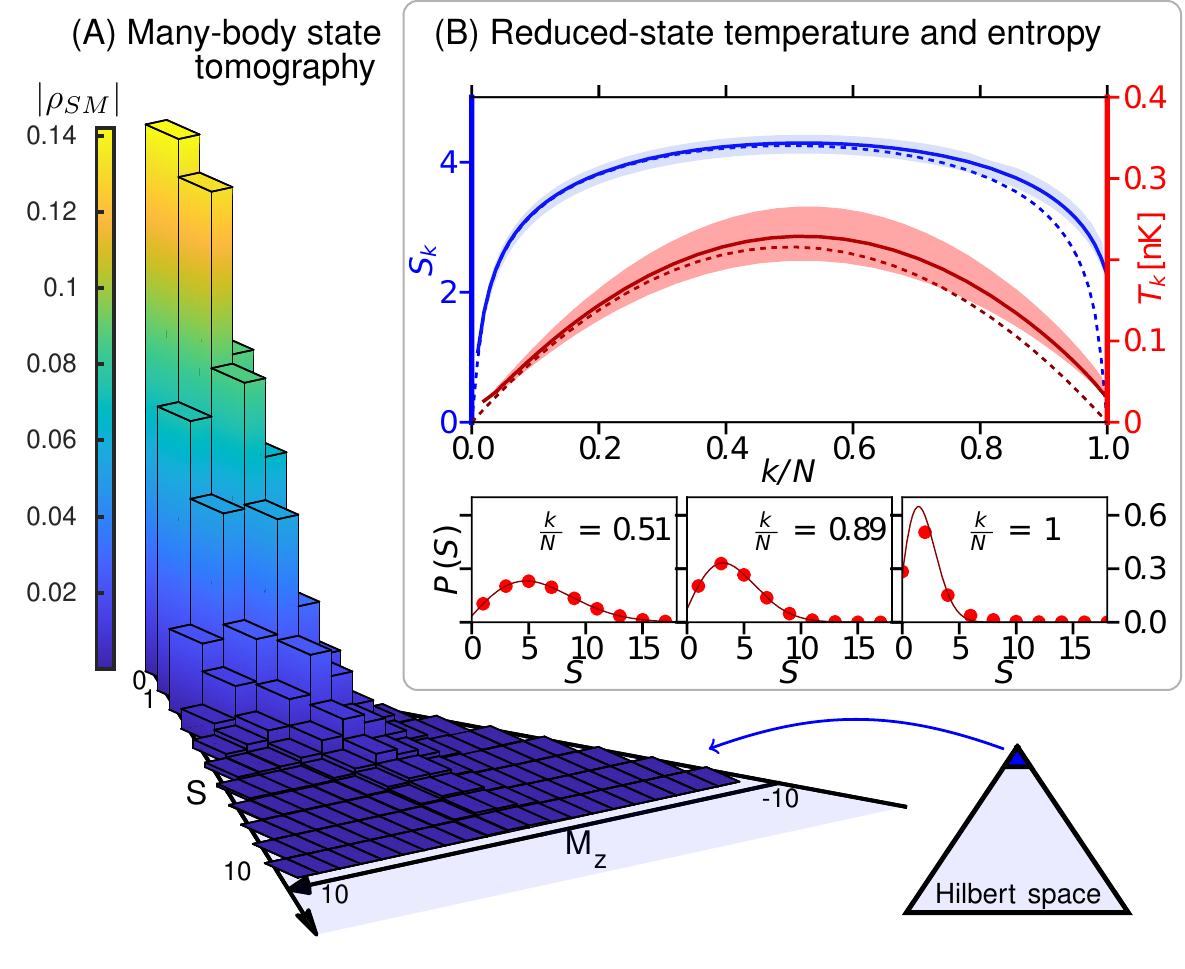}
	\caption{{\bf A.} Diagonal elements of the reconstructed state in the spin state basis. {\bf B.} The top panel shows the entropy (blue line) and temperature (red line) of the reduced density matrices. The shaded areas correspond to the $68\%$ confidence interval. The three lower panels show the measured spin distributions (red dots) and the thermal fits (solid lines) used to extract a temperature, for various values of $k/N$.}\label{figure4}
\end{figure}

We now turn to the tomography of the many-body state, which for a hundred atoms can be a formidable task \cite{dariano2003}. Here, we restrict to the spin state only, leading to a Hilbert space dimension $\sim10^4$. We know from our spin measurements that the state is very localized in the $|S,M\rangle$ basis, which allows for a faithful reconstruction using a moderate data set of $\sim 1100$ measurements after various spin rotations. We use the so-called ``Maximum-Likelihood" method, based on Bayesian inferences \cite{SM,dariano2003,lvovsky2004,strobel2014,peise2015}. The reconstructed density matrix $\rho^{(N)}$ is almost diagonal in the $|S,M\rangle$ basis and its diagonal elements are represented in Fig.\,\ref{figure4}{\bf A}. The lowest spin states, the singlet $S=0$ for $N$ even and $S=1$ for $N$ odd, are the most populated, and the first four spin manifolds contain $\simeq90\%$ of the total population. We recover a remarkable property of strongly correlated systems: while $\rho^{(1)}$ is completely mixed, the many-body density matrix $\rho^{(N)}$ has a very low entropy. 

For a complete picture, we computed the reduced spin density matrix $\rho^{(k)}$ associated to a subsystem of $k$ atoms, $\rho^{(k)}={\rm Tr}_{N-k}\rho^{(N)}$, where ${\rm Tr}_{N-k}$ designates the partial trace over the spin state of $N-k$ atom \cite{gessner2018}. Then we calculate the von Neumann entropy $S_k$ associated to $\rho^{(k)}$, as well as its temperature $T_k$ by fitting the spin distribution (Fig.\,\ref{figure4}\,{\bf B}). For the $N$-body singlet state, $\rho_{\rm s}^{(k)}$ corresponds to a thermal state at temperature \cite{SM}
\begin{align}
	T_k=\frac{U_s}{k_B}\frac{k}{N}\left(1-\frac{k}{N}\right)\,,
\end{align}
which is also plotted in Fig.\,\ref{figure4}\,{\bf B}. Measured and predicted quantities show the same qualitative behavior: For small $k$, the entropy increases, as for an uncorrelated system at non-zero temperature. Then, for $k$ slightly larger than $N/2$, we observe a back-bending of both entropy and temperature. This behavior can only be explained by the existence of entanglement between atoms \cite{vedral2002,islam2015}. For the whole system ($k=N$), 
the residual spin entropy $\sim 3.0$ agrees with the prediction of the stochastic $N$-body Schr\"odinger equation and can be attributed to the randomness of the parity of $N$ at the beginning of the evolution and to the residual decoherence during the preparation ramp \cite{SM}.

In conclusion, we prepared a quasi-pure many-body state of $N\sim 100$ atoms with spin 1, that corresponds to a fragmented condensate with 3 similarly populated single-particle states. This state
can be seen as the association of all but a few atoms into singlet pairs. Recently, a twin-Fock state of $N\sim 10^4$ atoms, which should ideally correspond to a $2$-fragment condensate, was analyzed by its number squeezing, a two-body observable \cite{luo2017}. 
Here we could reconstruct the full many-body state thanks to the use of a detection scheme at the single atom level. Our relatively small atom number allowed us to obtain a close-to-adiabatic following of the many-body ground state from an initially uncorrelated situation, even though the targeted state at zero magnetic field is the critical point of a quantum phase transition \cite{hoang2016}.  Performing a similar experiment in an optical lattice, where new phases are expected to emerge from the interplay between spatial and spin degrees of freedom,  constitutes an exciting direction for future work both for fundamental aspects \cite{yip2003,imambekov2004} and for applications in quantum metrology \cite{Urizar2013}.

\bibliographystyle{Science}
\bibliography{BiblioFragmentation}

\section*{Acknowledgments}
We thank the members of the BEC group at LKB for insightful discussions. This work was supported by ERC (Synergy Grant UQUAM). LKB is a member of the SIRTEQ network of R\'egion Ile-de-France.

\newpage
\subsection*{Methods}

\paragraph{Preparation of the initial state.}
The experimental sequence to produce the initial polar state  $|0\rangle^{\otimes N}$, with $N\simeq 100$, is composed of the following steps:
\begin{enumerate}
	\item We prepare a BEC in a shallow crossed optical dipole trap, in a bias magnetic field of $1\,$G. At this stage we have about $2000$ atoms, and $n_0\simeq 0.85$.
	\item We ``spin-distill" the state: atoms in the $m=\pm1$ states are removed by applying a magnetic force that pulls them out of the trap.
	\item Using a RF field resonant with the Zeeman splitting, we transfer most of the atoms in $m=\pm1$, and then spin-distill again. We are left with about $104 \pm 15$ (one standard deviation) atoms in $m=0$, and none detectable in $m=\pm1$.
	\item We recompress the trap. The trap frequencies at the end of the compression are about $(2.0,2.8,2.0)\,$kHz.
\end{enumerate}

\paragraph{Calibration of the interaction strength $U_s$.}
The interaction strength $U_s$ is calibrated looking at oscillations of the $m=\pm1$ populations after a quench of the magnetic field starting from the state $|0\rangle^{\otimes N}$. For ${U_s}/{N}\ll q$, the Bogoliubov approximation is valid and the number of atoms in $m=\pm1$ evolves as $N_{\pm 1}=({U_s}/{\hbar\omega_B})^2\sin^2(\omega_B t)$,
where $\hbar\omega_B=\sqrt{q(q+2U_s)}$. From a fit of $N_{\pm1}(t)$ with $U_s$ as the only free parameter, we extract $U_s/h = 18.8 \pm 2.4\,$Hz.

\paragraph{Adiabatic ramp to quasi-null magnetic field.}
To prepare the fragmented condensate, we use an adiabatic passage  driving the system from the uncorrelated state $|m=0\rangle_z^{\otimes N}$ at large $B$ to the targeted state at $B\approx 0$. The starting point corresponds to $B=1.000\,(4)\,$G, for which $q_i/h= 277\,$Hz is much larger than $U_s/h= 18.8\,$Hz. In $1\,$s we ramp down $B$ to $4.0\,(6)\,$mG ($q_f/h=4\,$mHz), achieving $q_fN^2/U_s \simeq 2.5$. For this $q_f$, the calculated ground state $|\Phi_f\rangle$ is fragmented, with an overlap of $92\,\%$ (resp. $99\,\%$) with the $S=0$ (resp. $S=1$) state for $N$ even (resp. odd). The optimal duration and final value of the ramp are set by a compromise between the adiabaticity requirement and decoherence, which we settle empirically. 
We aim for a minimal production of excitation distributed over the whole ramp for a given duration. We design the ramp such that the energy gap $\Delta E(t)$ between the ground state and the first excited state verifies
\begin{align}
\left|\frac{{\rm d} \,\Delta E}{{\rm d}t}\right|=\epsilon\,\Delta E^2\,, \label{eq.ramp}
\end{align}
where $\epsilon$ is a number that should be small in order to deviate only marginally from the adiabaticity.
For almost the whole ramp we have ${U_s}/{N^2}\ll q\ll U_s$, such that the Bogoliubov approximation applies and gives $\Delta E \propto B$, where $B$ is the magnetic field. Integration of Eq.(\ref{eq.ramp}) gives
\begin{align}
B(t)=\frac{B_i}{1+\frac{B_i t}{B_f t_f}}\,
\end{align}
where $B_i$ ($B_f$) is the initial (final) magnetic field and $t_f$ the duration of the ramp. Here we made the approximation $B_i\gg B_f$. 

\paragraph{Imaging.}
The three spin states are spatially separated using a Stern-Gerlach splitting during time of flight. We then shine an optical molasses and collect the fluorescence light. From this signal we extract the population of each Zeeman state. The experimental set-up, sequence and the image processing are described in detail in \cite{qu2020}. For the present experiment, the standard deviations of the atom number measured on empty images are $[\Delta N_+,\Delta N_0,\Delta N_-]=[1.2,1.4,1.1]$. The main source of noise is the  shot-noise on stray light. The inhomogeneity of the stray light background is responsible for the slightly different noise levels for the three Zeeman states. Note that the Stern-Gerlach separation requires a large magnetic field. Therefore, at the end of the adiabatic ramp, we quench the magnetic field to a value of $\sim 2.5$ G, such that $q\sim 1.7\,$kHz is much larger than $U_s$. In this condition, spin-mixing dynamics is frozen and the populations of the Zeeman states are conserved. We have verified this point both numerically and experimentally, by scanning the duration of the quench of the magnetic field. For short enough quenches, we did not observe any evolution of the Zeeman state populations. 

\paragraph{Measurement of $S_x$.}
The measurement of the populations of the Zeeman states gives access to the value of the spin along the magnetic field axis, $S_z=N_{+1}^{(z)}-N_{-1}^{(z)}$. In order to measure the other spin components, we use a resonant radio-frequency field to couple the Zeeman sublevels. In the rotating wave approximation (RWA), this results in a rotation in spin space, around the $y$-axis (determined by the phase of the RF field). For the RWA to be valid, the Larmor frequency $f_L$ must be much larger than the Rabi frequency. This is technically difficult to achieve at the final field of $4\,$mG, for which $f_L=2.8\,$kHz. To overcome this issue, we quickly (in $6\,$ms) ramp the magnetic field up to $50\,$mG. Then $f_L=35\,$kHz and $q/\hbar=0.7\,$Hz. Keeping a small $q$ is important in order to limit the evolution of the spin component $\hat{S}_\nu$ (${\rm i}\hbar \,{\rm d}\hat{S}_\nu/{\rm d}t=-q[\hat{S}_\nu,\hat{N}_0]$) until the rotation maps $\hat{S}_\nu$ onto ${\hat S}_z$. Then, we quench the field to $\sim 2.5\,$G so that ${\hat S}_z$ is conserved until the measurement. We verified numerically and experimentally (scanning the intermediate field and the duration of the ramp) that the small ramp of $q$ before the rotation did not lead to a detectable evolution of the spin.

\paragraph{Measurement of $S_y$.}
To perform a rotation around the $z$ axis, we simply add a delay before the RF pulse. Since the bias field is oriented along $z$, the spin then naturally precesses around this axis. We use this procedure to check the isotropy of the spin in the transverse $xy$ plane and to extract the single particle density matrix. 

\end{document}